# CHARACTERIZING DISEASES AND DISORDERS IN GAY USERS' TWEETS

**Frank Webb**
University of South Carolina
fwebb@email.sc.edu

**Amir Karami**
University of South Carolina
karami@sc.edu

**Vanessa Kitzie**
University of South Carolina
kitzie@mailbox.sc.edu

**ABSTRACT**

A lack of information exists about the health issues of lesbian, gay, bisexual, transgender, and queer (LGBTQ) people who are often excluded from national demographic assessments, health studies, and clinical trials. As a result, medical experts and researchers lack holistic understanding of the health disparities facing these populations. Fortunately, publicly available social media data such as Twitter data can be utilized to support the decisions of public health policy makers and managers with respect to LGBTQ people. This research employs a computational approach to collect tweets from gay users on health-related topics and model these topics. To determine the nature of health-related information shared by men who have sex with men on Twitter, we collected thousands of tweets from 177 active users. We sampled these tweets using a framework that can be applied to other LGBTQ sub-populations in future research. We found 11 diseases in 7 categories based on ICD 10 that are in line with the published studies and official reports.

**Keywords**

Gay, LGBTQ, social media Twitter, text mining, health

## INTRODUCTION

Medical researchers and practitioners have increased attention to the health issues of lesbian, gay, bisexual, transgender, and queer (LGBTQ) people. One reason for this increased attention is that LGBTQ populations experience health disparities compared to straight and cisgender people, including higher rates of HIV and depression, and lower rates of preventative care (Gonzales, G., Przedworski, J., and Henning-Smith, C., 2016). Prior research data are limited to small samples that render it difficult to generalize the results to the world population. One reason for this limitation is the lack of LGBTQ representation in national health studies; only recently has a US federal initiative added questions for sexual orientation and gender identity. Since there exists no consistent health data collection policy, results from prior research cannot be directly compared.

The National Health Interview Survey (NHIS) identified that 3.4% of US adults are non-straight people including 1.6% gay people. The lack of information about LGBTQ health issues poses a significant threat to their health outcomes since it limits the responses of medical professionals to their unique disparities (Sell, R. L., and Holliday, M. L., 2014). Even removing structural limitations related to LGBTQ representation in national data collection, providing a good sample of LGBTQ populations is not easy. LGBTQ people comprise a relatively small proportion of the world population. Additionally, sub-groups within LGBTQ populations, such as gay men, experience unique health disparities relative to the LGBTQ population at large (Graham, R., Berkowitz, B., Blum, R., Bockting, W., Bradford, J., de Vries, B., Garofalo, R., Herek, G., Howell, E., Kasprzyk, D. and Makadon, H., 2011). Although several LGBTQ health-related studies have been developed, this research area needs more research and perspectives to detect the unique needs of LGBTQ people.

Millions of users share their stories and life experiences on social media every day. For example, people post 500 million tweets per day on Twitter covering different topics from health to politics. Social media platforms play a crucial role in the lives of LGBTQ people, who can transcend physical barriers such as geography to connect with similar others to share their collective concerns and search for their information in a safe space. (Byron, P., Rasmussen, S., Wright Toussaint, D., Lobo, R., Robinson, K. H., and Paradise, B., 2017). Privacy and stigma pose important barriers to LGBTQ people sharing their sexualities and gender identities in social media (Byron, P., Albury, K., and Evers, C., 2013). The accessible data in social media is a great opportunity to track LGBTQ health concerns. Among different social media platforms, Twitter has facilitated data collection using Application Programming Interfaces (API). This new data collection method is an efficient approach to minimize cost and time. Several studies have been developed based on using Twitter data in different domains such as health (Shaw, G., and Karami, A., 2017), election analysis (Karami, A., Bennett, L. S., and He, X., 2018), psychology (Golder, S. A., and Macy, M. W., 2011) and disaster management (Cameron, M. A., Power, R., Robinson, B., and Yin, J., 2012). Few studies examine how LGBTQ people share their health issues using social media. This research focuses on gay men. This paper proposes a framework to collect and analyze the tweets of individuals self-identifying as gay to characterize health issues experienced by this group of users. There are different health issues, but this study focuses on diseases and disorders in the collected tweets. This paper provides a new direction at the intersection of LGBTQ, health, medical, and data science.

**RELATED WORKS**

The United States Office of Disease Prevention and Health Promotion's *Healthy People 2020* initiative identified LGBTQ people as facing significant health disparities including higher rates of substance abuse, mental illness and suicide attempts, and youth homelessness, as well as lower rates of health care utilization. Gay men face the greatest risk of HIV and other STIs, as well as high rates of recreational drug use (Office of Disease Prevention and Health Promotion, 2017). These disparities are produced, in part, to the discriminatory treatment LGBTQ people face when receiving health care. Health care providers receive a cumulative median of five hours of LGBTQ-related curricular content during the whole of their medical instruction, and prior research denotes that medical professionals have an implicit preference for heterosexual patients (Obedin-Maliver J, Goldsmith ES, Stewart L, & et al, 2011; Sabin, Riskind, and Nosek, 2015). Together, these sparse instructions and biases among medical professionals can lead to negative experiences for gay men with their healthcare providers, discouraging them from disclosing their gay identities and decreasing their utilization of health care resources.

*Healthy People 2020* also identifies the lack of sexual orientation and gender identity information associated with medical records as an area of improvement to lessen disparities for LGBTQ populations. As of 2014, only eight national data systems included information on lesbian, gay, and bisexual people (Office of Disease Prevention and Health Promotion, 2017). Efforts to collect the appropriate data focus on standardized terminology, confidentiality, and clarity of questions, as well as reducing barriers to patients' willingness to offer the information. Because sexual orientation and behavior do not necessarily correspond, heterosexual-identifying patients may not be willing to report having had a same-sex partner, limiting the providers' knowledge for treatment. By creating comprehensive records of orientation and behavior associated with patients' medical records, better education and treatment methods can be utilized to reduce disparities (Institute of Medicine (US) Board on the Health of Select Populations, 2013).

The first step in the public health cycle is assessment with two main activities: monitor health status and diagnose health problems/hazard. The next step, policy development, depends on the first step (Harrell, J. A., and Baker, E. L., 1994). However, it is difficult to track and detect health disparities among LGBTQ populations due to the limited nature of the data available. Social media have provided a great platform for users to share their concerns with respect to different issues like health. This new opportunity can open a new research direction for the experts in health and medical informatics with the purpose of LGBTQ health surveillance.

Among different social media, Twitter allows users to publicly share short-form text content online, creating an opportunity for researchers to analyze the collective opinions, thoughts, and concerns of communities in near real-time in a way traditional methods cannot. The text-based data collected from Twitter can be analyzed using data mining methods. Twitter data has been used for different health applications such as tracking influenza (Signorini, A., Segre, A. M., & Polgreen, P. M., 2011), infectious intestinal disease (Zou, B., Lampos, V., Gorton, R., and Cox, I. J., 2016), migraine headache (Nascimento, Thiago D., Marcos F. DosSantos, Theodora Danciu, Misty DeBoer, Hendrik van Holsbeeck, Sarah R. Lucas, Christine Aiello, Leen Khatib, and MaryCatherine A. Bender., 2014), and diabetes, diet, exercise, and obesity (Abbar, S., Mejova, Y., and Weber, I., 2015; Karami, A., Dahl, A. A., Turner-McGrievy, G., Kharrazi, H., and Shaw, G., 2018).

Twitter data can be used to supplement health knowledge because the users are more willing to share some information online than they would with health care providers. Previous studies have dealt with analyzing different health issues without a focus on LGBTQ people's health issues. This study investigates gay men's activity on Twitter to better understand the common diseases and disorders they face and to inform a methodological framework that can be used to collect data from other LGBTQ sub-populations.

**METHODOLOGY & RESULTS**

Our approach uses computational methods to collect data and analyze tweets for disclosing characteristics of opinions in the tweets of gay users with respect to diseases. This study has three phases: data collection, health-related tweets detection, topic discovery and analysis.

**Data Collection**

Gay Twitter users were identified as part of a larger LGBTQ user identification process. Users were initially collected by manually checking the followers of prominent LGBTQ users and organizations. These users were selected based on self-identification (disclosure) as part of the LGBTQ community in their profile and for their activity on Twitter being primarily personal posts documenting their lives. For orientation confirmation, a variety of factors were considered. Beyond explicit identifying statements, the use of gay terminology for self-descriptions and interactions with gay content were considered for categorizing the user as a gay man. Posts using terms such as "Grindr", "top", "bottom", "twink", and "bear" were indicators of the user's knowledge of gay communities and a willingness to publicly discuss those topics. Additionally, liking, retweeting, and replying to similar content was also considered. While there is the possibility of these men being bisexual and just having more prominent same-sex content, the lack of indications for heterosexuality made improper categorization unlikely. These characteristics helped us find 177 active gay users in Twitter. In the next step, we used the Twitter API in R platform to collect up to 3200 tweets (Twitter API limitation) for each of those 177 users for 487,155 tweets in total.

**Health-Related Tweets Detection**

To focus on just personal opinions and experiences, we removed retweets and the tweets containing URLs. We applied the Linguistic Inquiry and Word Count (LIWC) dictionary to these remaining tweets. LIWC captures thoughts, feelings, personality, and motivations. It has a health-related dictionary that detects health-related words in a corpus. From our sample of unique tweets, we extracted those containing at least one health-related word, yielding 4,016 tweets. Table 1 displays a sample of the collected health-related tweets about insomnia, HIV, stress, and cancer.

| Tweet 1 | *i got insomnia* |
| Tweet 2 | *living with #HIV, is addressing the convention. Bravo,* |
| Tweet 3 | *Stressing about life and rushing an English paper that's due at midnight even …* |
| Tweet 4 | *also this is standard cancer checkup stuff so no need to be alarmed* |

**Table 1. Sample of Health-related Tweets**

**Topic Discovery and Analysis**

The huge amount of text data in different applications has created a need for text mining to detect interesting pattern**.** Text mining has been used for different applications such as spam detection (Karami A. and Zhou L., 2014; Karami, A. and Zhou, B., 2015), and health corpora analysis (Karami, A., Gangopadhyay, A., Zhou, B., and Kharrazi, H., 2015a)**.** We employed a well-known text mining method, Latent Dirichlet Allocation (LDA), to describe the 4016 health-related tweets. As a topic modeling method, LDA clusters semantically related words such as "human recourse", "marketing", and "finance" into a topic whose theme is "business" (Karami, A., 2015). This model is based an assumption that if there are multiple topics in a corpus, LDA assigns a degree of membership for each word in the corpus with respect to each of the topics (Karami, A., Gangopadhyay, A., Zhou, B., & Karrazi, H., 2015b). Each topic is a distribution of semantically related words (Karami, A., Gangopadhyay, A., Zhou, B., and Kharrazi, H., 2017).

**Results**

Then we analyzed the content of each of the topics to find the overall theme for each of them. For example, we assigned "addiction" label to a topic having these words: *mental, health, illness, control, and addictions* (Table 2). From this analysis, we derived 11 topics. We then categorized these 11 detected topics based on ICD 10 (International Classification of Diseases) classification[1]. Our results show that gay Twitter users have discussed about 7 categories including 11 diseases:

---

[1] http://apps.who.int/classifications/icd10/browse/2016/en#/G43

- Endocrine, nutritional and metabolic diseases: **obesity**
- Diseases of the circulatory system: **heart attack**
- Diseases of the nervous system: **insomnia**, **seizures**, and **stroke**
- Infectious and parasitic diseases: **HIV**
- Neoplasms: **cancer**
- Diseases of the respiratory system: **flu** and **asthma**
- Mental and behavioral disorders: **stress** and **addiction**

| Obesity/Heart Attack | Insomnia/ HIV | Seizures | Stroke | Cancer |
|---|---|---|---|---|
| heart | Work | alcohol | violence | living |
| attack | week | afford | clinical | suffer |
| obesity | addiction | infection | tweet | vomit |
| stomach | insomnia | seizures | stroke | cancer |
| lowcarb | hiv/#aids | woke | gained | experience |
| Flu | Asthma | Stress | Addiction | HIV |
| live | plan | shit | mental | *allergies* |
| living | obamacare | horrible | health | *hiv/aids* |
| days | access | drink | illness | *tumor* |
| flu | asthma | stress | control | *extreme* |
| find | healthcare | headaches | addictions | *disease* |

**Table 2. A Sample of Topics**

Our results are in line with formal and published reports including HIV/AIDS (CDC, 2014), stress and insomnia (Berg, M. B., Mimiaga, M. J., and Safren, S. A., 2008), stroke and seizures (Rosendale, N., and Josephson, S. A., 2015), cancer, asthma, and heart attack (Gonzales, G., Przedworski, J., and Henning-Smith, C., 2016), obesity and cancer (Wender, R., Sharpe, K. B., Westmaas, J. L., and Patel, A. V., 2016), addiction (Kecojevic, A., Wong, C. F., Schrager, S. M., Silva, K., Bloom, J. J., Iverson, E., and Lankenau, S. E., 2012), and flu (Whitehead, J., Shaver, J., and Stephenson, R., 2016). Moreover, the detected associations have also addresses in the literature including obesity and heart attack (Eckel, R. H., 1997), insomnia and HIV (Buchanan, D. T., McCurry, S. M., Eilers, K., Applin, S., Williams, E. T., and Voss, J. G., 2016), seizures and infection (Wong, M. C., and Labar, D. R., 1990), stroke and violence (Paradiso, S., Robinson, R. G., and Arndt, S., 1996), cancer and vomiting (Matsha, T., Stepien, A., Blanco-Blanco, E., Brink, L. T., Lombard, C. J., Van Rensburg, S., and Erasmus, R. T., 2006), stress and headache (Lipton, R. B., Stewart, W. F., and Liberman, J. N., 2002), mental health and addiction (Lasser, K., Boyd, J. W., Woolhandler, S., Himmelstein, D. U., McCormick, D., and Bor, D. H., 2000), and HIV and allergy (Pirmohamed, M., & Park, B. K., 2001). It seems that the relation between Asthma and Obamacare is based on the concern of users with respect to covering the cost of Asthma in Obamacare plan.

**CONCLUSION**

LGBTQ people experience more health problems than other people; however, the biggest health challenge to these people is the lack of information about the health issues. Moreover, the current available data and data collection methods have some limitations. This study proposes the first step in developing a computational approach to collect, analyze, and represent common diseases on Twitter posts of gay users. This approach detects health-related posts and discovers disease-related topics in thousands of tweets. With billions of social media users, this research opens a new interdisciplinary direction at the intersection of data science, health, informatics, and social media.

This paper has found hundreds of self-reported gay people in Twitter and collected thousands of tweets. Two text analysis approaches have been applied on the tweets to find health-related tweets and the embedded topics. Then, the 11 disease-related topics were detected, labeled, and assigned to 7 major categories based on ICD 10. The proposed approach and results can be used to support public health experts and policy makers in tracking and understanding diseases with respect to LGBTQ people. With further work to increase the sample population size and incorporating temporal and spatial data, this model can be used to produce targeted interventions to help minimize the health disparities present in gay men's populations.

**ACKNOWLEDGMENTS**

This research is supported in part by the University of South Carolina Honors College Exploration Scholars and Magellan Scholar Programs. All opinions, findings, conclusions, and recommendations in this paper are those of the authors and do not necessarily reflect the views of the funding agency.